\documentclass{elsart}

\usepackage{natbib}

 \usepackage{graphics}
 \usepackage{graphicx}
 \usepackage{epsfig}

\usepackage{amssymb}

\newcommand{\apj}{ApJ}
\newcommand{\apjl}{ApJL}
\newcommand{\mnras}{MNRAS}
\newcommand{\aap}{A\&A}

\begin{document}

\begin{frontmatter}



\title{Modeling Pulsar Wind Nebulae}


\author{Bucciantini, N.}

\address{Astronomy Department, University of California at Berkeley, 601 Camblell Hall, Berkeley, CA, 94706, USA, Email: nbucciantini@astro.berkeley.edu}

\begin{abstract}

In the last few years, new observations by CHANDRA and XMM have shown that Pulsar Wind Nebulae present a complex but similar inner feature, with the presence of axisymmetric rings and jets, which is generally referred as {\it jet-torus structure}. Due to the rapid growth in accuracy and robustness of numerical schemes for relativistic fluid-dynamics, it is now possible to model the flow and magnetic structure of the relativistic plasma responsible for the emission. Recent results have clarified how the jet and rings are formed, suggesting that the morphology is strongly related to the wind properties, so that, in principle, it is possible to infer the conditions in the unshocked wind from the nebular emission. I will review here the current status in the modeling of Pulsar Wind Nebulae, and, in particular, how numerical simulations have increased our understanding of the flow structure, observed emission, polarization and spectral properties. I will also point to possible future developments of the present models.

\end{abstract}


\end{frontmatter}



\section{Introduction}
\label{sec:intro}

Pulsar Wind Nebulae (PWNe) are bubbles of relativistic particles and magnetic field produced by the interaction of the ultra-relativistic pulsar wind with the ambient medium (ISM or SNR). The best example of PWN and the prototype of the entire class is the Crab Nebula. The basic model of a PWN was first presented by \citet{ree74}, and developed in more detail by  \citet{ken84a,ken84b} (KC84 hereafter), and \citet{emm87}. The ultra-relativistic wind produced by the pulsar is confined inside the SNR, and slowed down to non relativistic speeds in a strong termination shock (TS). At the shock the toroidal magnetic field of the wind is compressed, the plasma is heated and particles are accelerated to high energy. A bubble of high energy particles and magnetic field is produced where the post-shock flow  expands at a non relativistic speed toward the edge of the nebula. This model was particularly successful and was able to explain many of the observed properties of the Crab Nebula. The continuous, non-thermal, very broad-band spectrum, extending from Radio to X-ray, with spectral index in the range 0-1.2, steepening with increasing frequencies, was modeled as synchrotron emission, associated with the gyration of accelerated high energy particles in the compressed magnetic field \citep{ver93,ban99,wei00,wil01,mor04}. The presence of an under-luminous region, centered on the location of the pulsar, was interpreted as evidence for the ultra-relativistic unshocked wind. Polarization measures \citep{wil72,vel85,sch79,hic90,mic91} show that emission is highly polarized and the nebular magnetic field is mostly toroidal, as one would expect from the compression of the pulsar wind. Interestingly this model clearly explains why Crab Nebula appears bigger at smaller frequencies \citep{ver93,bie97,ban98}. High energy X-ray emitting particles have a short lifetime for synchrotron losses, and they are not advected far away from the TS; in contrast, the synchrotron lifetime for radio particles is longer than the age of the nebula, so they fill the entire volume. Based on this simple model it is possible to constrain some of the properties of the pulsar wind, at least at the distance of the TS. To explain the dynamics of the plasma, as well as the emission properties of the nebula, the Lorentz factor of the wind is estimated to be $\sim 10^6$, and the ratio between Poynting flux and kinetic energy $\sigma\sim 0.003$. The model also explains why many PWNe ({\it i.e.} Crab Nebula, 3C58) have an elongated axisymmetric shape: the toroidal magnetic field in the nebula exerts a higher pressure on the confining ambient medium along the polar direction than on the equatorial plane \citep{beg92,van03}.

In the model by KC84 the SNR has only a passive role providing the confinement of the PWN. However, the details of the flow structure in the PWN and, as a consequence, of the emission properties depend critically on the boundary condition with the SNR. Given the complexity of the interaction and the existences of different timescales, a detailed study of the evolution of the system PWN-SNR, has been possible only recently, thanks mostly to the improvement in computational resources \citep{van01,blo01,me03,van04}. Comparing the energy in the SNR ($\sim 10^{51}$ ergs) to the total energy injected by the pulsar in the PWN ($\sim 10^{49}$ ergs), it is immediately evident that the PWN cannot significantly affect the global evolution of the SNR, but on the other hand, the evolution of the SNR can have important consequences for the PWN. Three main phases can be identified in the evolution of a PWN inside a SNR (for a more complete discussion of PWN-SNR evolution see \citet{ren84} and \citet{gae06}). In the first phase the PWN expands inside the cold SN ejecta. The SN ejecta are in free expansion, so this phase is referred as {\it free expansion phase}. During this phase the pulsar spin-down luminosity is high and almost constant, and the nebula can be easily observed in high energy emission. The Crab Nebula is presently supposed to be in this phase. The free-expansion phase terminates after about 1000-3000 yr, when the PWN reaches the reverse shock in the SNR shell. From this moment, the evolution of the PWN is deeply modified by interaction with the more massive and energetic SNR ejecta. In the absence of a central PWN, the reverse shock is supposed to reach the center of the SNR in about 5000-10000 yr. As the SNR ejecta compress back the PWN, its pressure increases. The compressed PWN will then push back the inner edge of the SNR. This phase is called ``reverberation phase'': the PWN might undergo several compressions and rarefactions. Even if the energy injection from the pulsar is negligible at these later times, PWNe can still be observed, due to the re-energization during compressions. Multidimensional studies have shown that the SNR-PWN interface is Rayleigh-Taylor unstable during compressions \citep{blo01}. This can cause efficient mixing of the pulsar wind material with the SNR, and ultimately with the ISM. The third phase corresponds to the Sedov phase of the SNR: the PWN expands adiabatically inside the heated SNR. Given the absence of  energy injection, a PWN in this stage is probably only observable as a faint extended radio source. Recent studies have shown that this simple evolutionary picture can be further complicated if one takes into account the pulsar kick velocity with respect to the center of the SNR \citep{van03b,van04,me05}. If the kick is strong enough, PWNe can be revealed also at later time as bow-shock nebulae \citep{me05}, when the pulsar emerges from the SNR, and interacts directly with the ISM.  

\begin{figure}
\label{fig:1}
\resizebox{\hsize}{!}{
\includegraphics{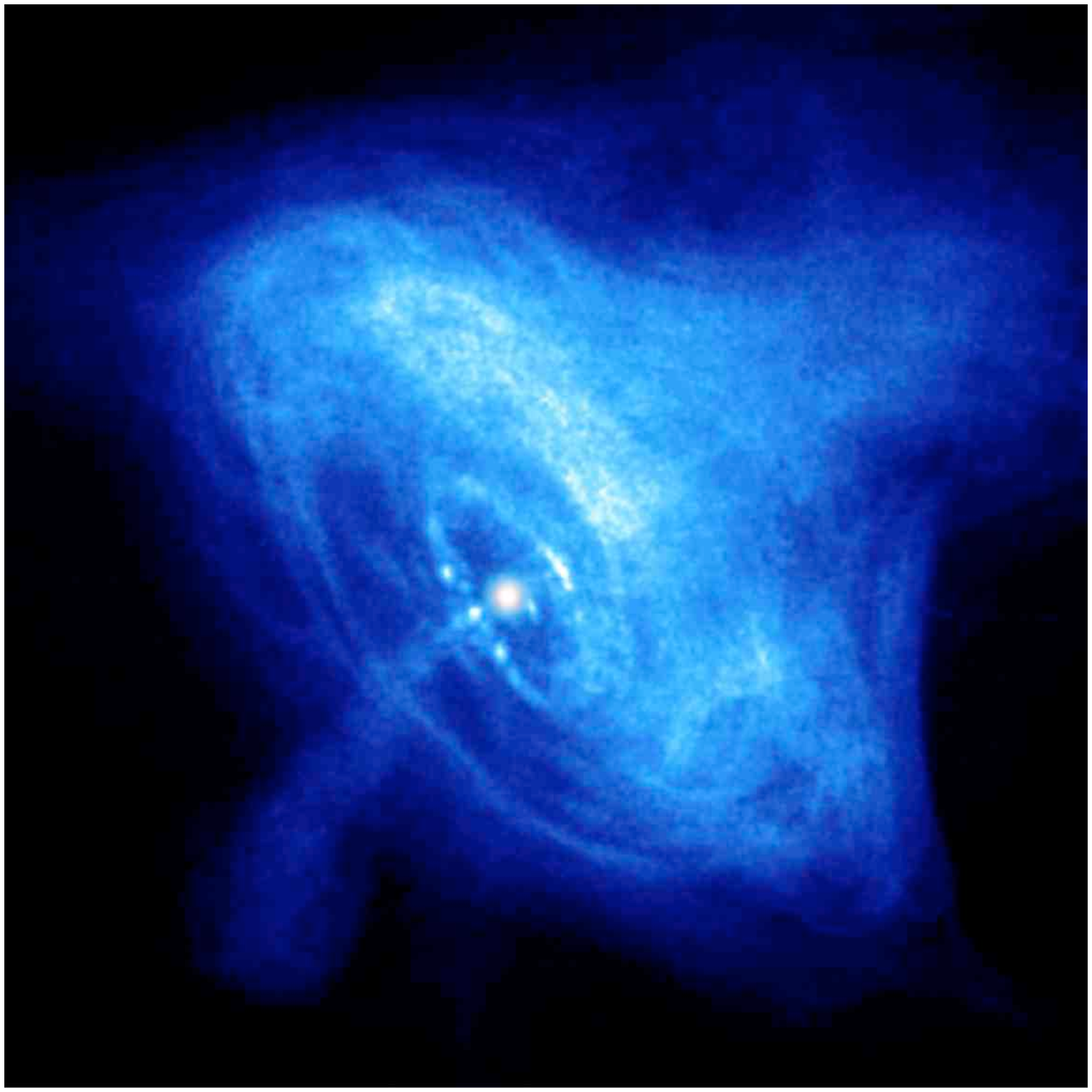}\hspace{0.2truecm}\includegraphics[scale=1.1]{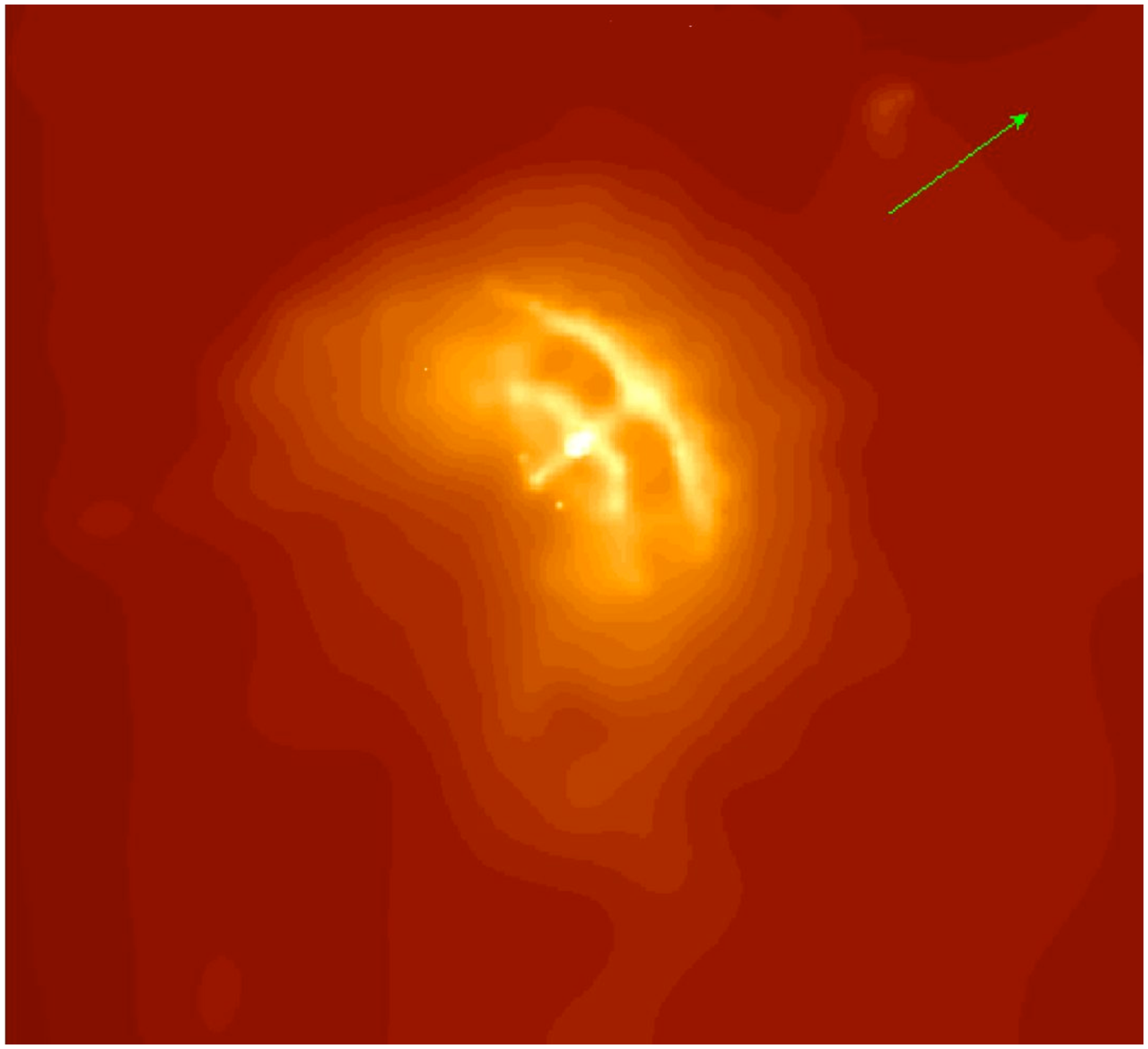}\hspace{0.2truecm}\includegraphics[scale=0.99]{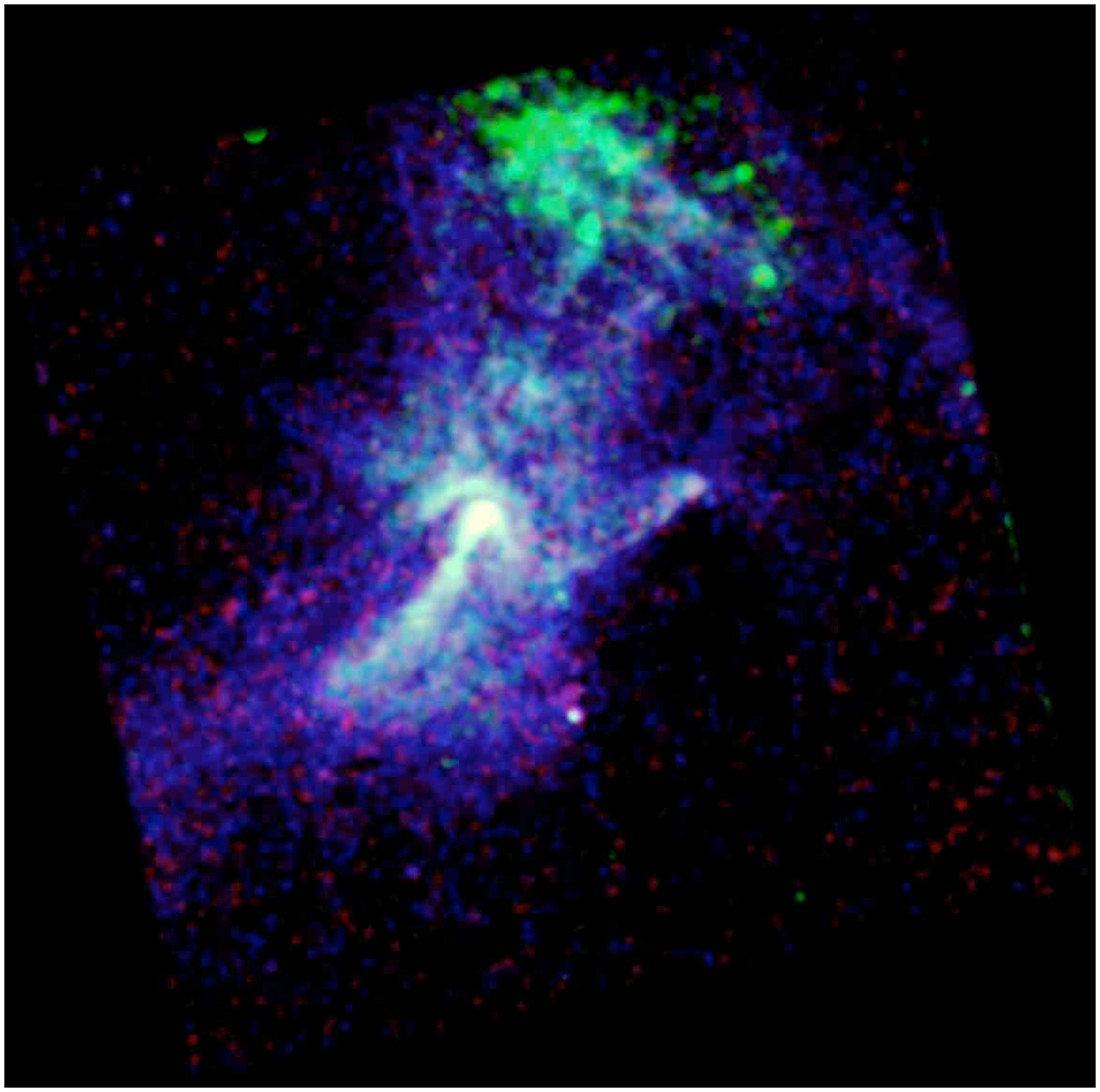}}
\caption{X-ray CHANDRA images (NASA/CXC/SAO) of the inner region of PWNe: {\it jet-torus} structure. Left: Crab Nebula. Center: Vela. Right: B1509}
\end{figure}

The interest of the scientific community for PWNe has received a considerable impulse from recent optical and X-ray images from HST, CHANDRA and XMM-Newton. These new data show that the inner region of PWNe is characterized by a complex axisymmetric structure, generally referred as {\it jet-torus structure} (Fig.~\ref{fig:1}). First observed in the Crab \citep{hes95,wei00}, it has been detected in many other PWNe \citep{pav03,gae02,lu02,rom03,sla04,cam04,rom05}. It consists of a main emission torus, in what is thought to be the equatorial plane of the pulsar rotation, often associated with multiple arcs or rings, a central knot, located in the vicinity of the pulsar, and one or two opposite jets along the polar axis, apparently originating very close to the location of the pulsar itself. It was immediately clear that the simple 1D radial model by KC84 was completely unable to explain the observed features. Even if qualitatively the existence of a main torus (the first feature of the jet-torus structure to be detected in the Crab Nebula) could be explained by assuming that the energy injection, or the particles acceleration efficiency were higher in the equatorial plane \citep{bog02a}, it was not possible to reproduce the observed luminosity by using the KC84 model. \citet{shi03} was the first to realize 
 that, to explain the difference in brightness between the front and back sides of the torus in the Crab Nebula, the post-shock flow velocity should have been $\sim 0.3-0.4 c$, much higher than typical values for subsonic expanding flows. As the quality of data increased the inconsistency of the old model was made more clear. The existence of an inner ring, separated from the torus by an under-luminous region, was incompatible with the assumption of a smooth flow from the TS to the edge of the nebula. To justify the presence of the luminous X-ray knot inside the under-luminous supersonic wind region, dissipation of the magnetic field was thought to be at work, as a way to accelerate high energy particle in the supersonic wind. The presence of a jet was perhaps the most interesting feature for the theoretical issues it raised \citep{lyu01}. The fact that the jet originated close to the pulsar, suggested the possibility of  collimation of the pulsar wind. However, theoretical \citep{beg94,bes98} and numerical \citep{con99,bog01,gru05,kom06,me06} studies of relativistic winds from pulsars, failed to achieve any form of collimated energetic outflow. Finally photon index maps of Crab Nebula \citep{mor04}, show that the spectrum  flattens moving from the inner ring toward the main torus, while the standard model predicts a steepening due to synchrotron losses. Interestingly the symmetry axis of the jet-torus  appears to correspond to the major axis of the nebula, suggesting that the toroidal magnetic field is a key element in shaping the inner flow.

Only recently a theoretical understanding of a possible mechanism to explain the jet-torus structure has been presented, and it was soon recognized that magnetization and energy distribution in the pulsar wind were key elements. While the old KC84 model assumed an isotropic energy flux in the wind, it was known for a long time that the asymptotic solution by \citet{mic73},  recently confirmed with numerical simulations \citep{bog01,kom06,me06}, predicted a higher equatorial flux. An axisymmetric energy flux enhanced at the equator, produces an oblate TS with a cusp in the polar region \citep{bog02a,bog02b}. Given that the TS is oblique to the wind direction at higher latitudes, the post shock flow in the nebula can have speeds $\sim 0.3-0.5 c$. Moreover if hoop-stresses are at work in the mildly relativistic flow, jet collimation might occur in the post shock region \citep{lyu02,kan03}. The evident complexity of this scenario clearly prevented any sophisticated theoretical model, however, the recent development of more efficient and robust  numerical schemes for relativistic MHD \citep{kom99,ldz03,gam03}, has open the possibility for a detailed numerical description. The initial numerical results \citep{kom04} show that, based on the theoretical assumption of axisymmetric energy flux and post-shock hoop stresses, it is indeed possible to reproduce the observed features. Subsequent studies have tried to refine this picture, by investigating if and how the emission features could be used to understand the conditions of the pre-shock wind, trying to constrain the nebular emission properties, polarization and spectral variations. Despite the success in explaining the formation of the jet-torus structure, the recent developments have also posed other questions, regarding the fine details and time variability, which have not yet been satisfactory answered. 


\section{Global structure  of PWNe}
\label{sec:glob}

\begin{figure}
\label{fig:2}
\resizebox{\hsize}{!}{
\includegraphics[scale=0.7]{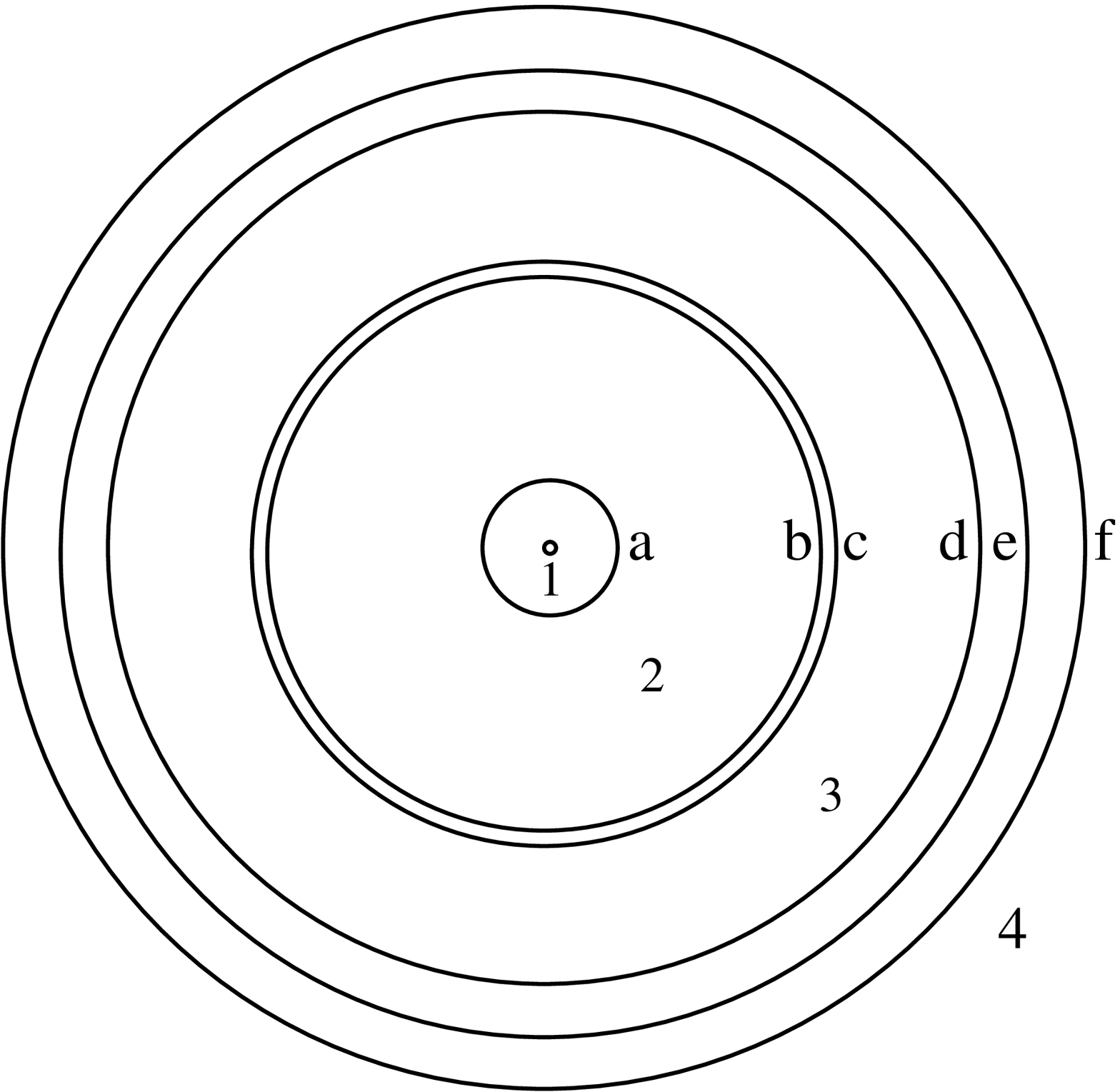}\hspace{0.5truecm}\includegraphics{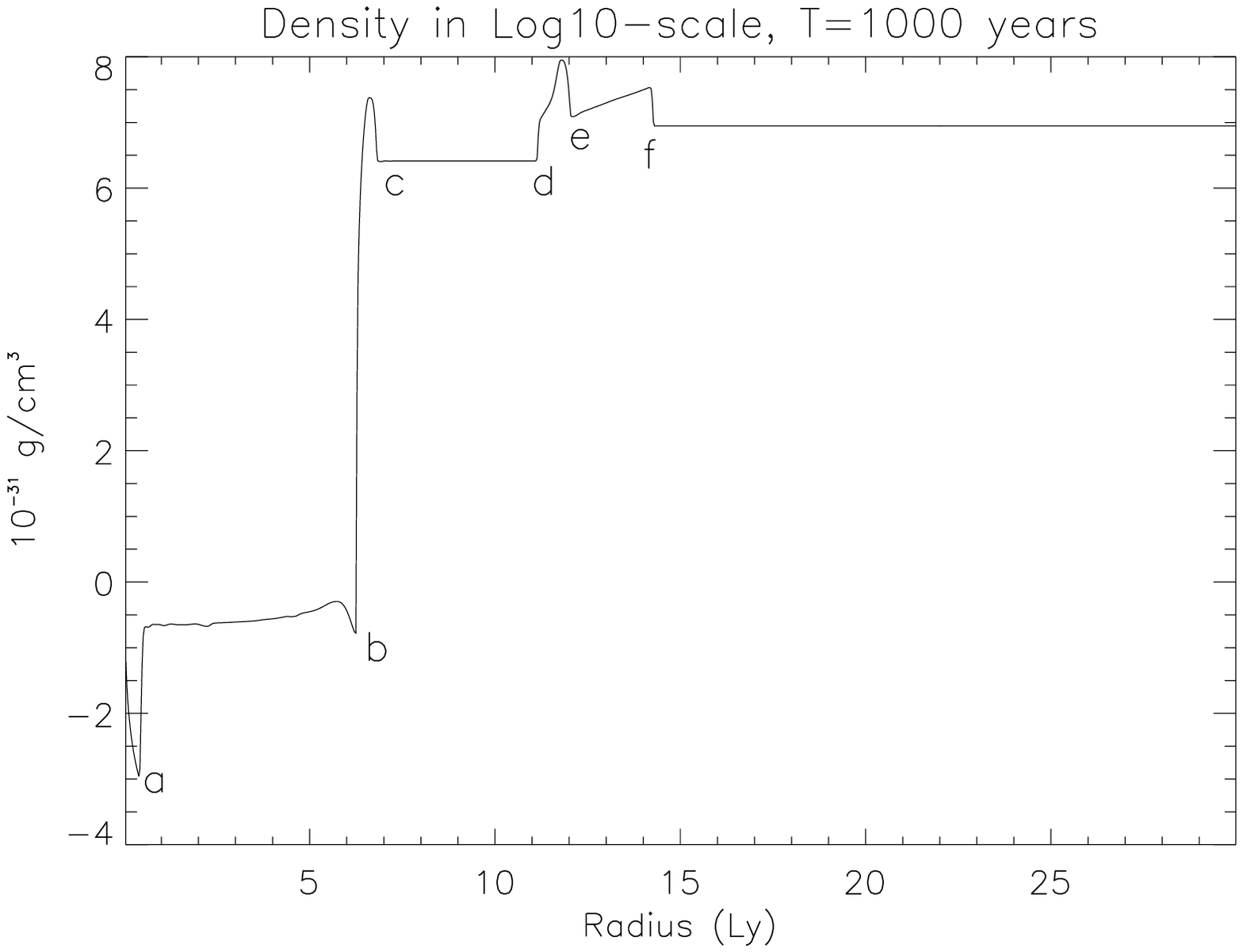}}
\caption{Left picture: schematic representation of the  global structure of the PWN in the first phase of its evolution inside a SNR. Right picture: density profile obtained from a time dependent numerical simulation in the relativistic hydrodynamical regime \citep{me03}. From the center the various regions are: 1- the  relativistic pulsar wind, 2- the hot magnetized bubble responsible for the non thermal emission, 3- the free expanding ejecta of the SNR, 4- the ISM. These regions are separated by discontinuities: a- the wind termination shock, b- the contact discontinuity between the hot shocked pulsar material and the swept-up SNR ejecta, c-the front shock of the thin shell expanding into the ejecta, d- the reverse shock of the SNR, e- the contact discontinuity separating the ejecta material from the compressed ISM, f- the forward SNR shock.}
\end{figure}

Let us first consider the global structure and morphology of PWNe, in the first phase of the PWN-SNR evolution, when the pulsar energy injection is still high and the PWN expands inside the free-expanding ejecta of the SNR. This corresponds to the present phase of the Crab Nebula, and in general is the phase in which X-ray emission from PWNe is expected to be stronger. In Fig.~\ref{fig:2} we present a simplified picture of a spherically symmetric PWN together with the result of a relativistic hydrodynamical simulation by \citet{me03}. Around the pulsar there is a region occupied be the ultra-relativistic pulsar wind. The wind is terminated in a strong shock. Downstream of the shock, there is a region filled with a low density hot plasma. The observed non-thermal emission comes from this region, which constitutes the PWN itself. As this region expands inside the high density, cold, supersonic ejecta of the SNR, a thin shell of compressed material is formed. The system is confined inside the SNR shell where the reverse shock, the forward shock, and the contact discontinuity, separating matter of the SN from the shocked ISM, are visible. In the following I will consider only the case of a pulsar at rest in the center of the system. The expansion velocity of PWNe in this early phase is known to be few thousands kilometer per second, while typical pulsar velocities are in the range 50-300 km/s. The pulsar kick will not appreciably modify the evolution and properties of the PWN before the onset of the reverberation stage. However, after the onset of the reverberation phase, major deviations in the global morphology and evolution of the PWN-SNR system can be expected (for a more complete study of the effects of pulsar velocity on the SNR-PWN system see \citet{van04}).

Given that the edge of the PWN is bounded by a compressed thin shell it is possible to derive analytically the evolution of the PWN radius with time, if one assume the SNR ejecta to be in a self-similar configuration. It turns out, in this case, that the evolution of the PWN is self-similar too, depending only on the pulsar luminosity, the density and velocity of the ejecta \citep{ren84,van01}. Even if analytic models assume a uniform pressure in the PWN, an assumption valid only in the pure hydrodynamical regime, simulations \citep{me03} have shown, in the 1D radial case, that the boundary pressure at the edge of the PWN does not depend on the magnetization, but only on the total energy injected by the pulsar. This suggests that global evolution of the size of PWN is not sensitive to the overall magnetization, thus strengthening the reliability of pressure driven thin shell models (Gelfand et al. these proceedings). The effect of spin-down and the drop in pulsar luminosity have also been studied \citep{me04}. Again, results show that the global evolution can be described analytically, using the thin shell approximation, and considering only the total energy injected by the pulsar. Simulations suggest however that a strong spin-down might change the ratio between the radius of the TS and that of the nebula. Given that such ratio has been used in the past to infer the magnetization of the wind, results show that temporal evolution in the pulsar properties can significantly alter the inferred values.  

The results of both analytic theory and spherically symmetric numerical simulations, show that the thin shell surrounding the PWN accelerates. Given that the density of the shell is much higher than the inertia of the relativistic plasma, the shell will be subject to Rayleigh-Taylor instability. The filamentary network of the Crab Nebula, observable in line emission, is supposed to be the result of such instability \citep{hes96}. Various simulations have investigated its growth \citep{jun98,me04b}, showing that  if the field is highly ordered at the boundary it can easily suppress the growth of parallel modes \citep{me04b}. Recent results (Stone, private communication) suggest that a more complex geometry of the magnetic field in the outer part of the nebula is required to reproduce the observed filamentary shape.

Magnetization plays a key role also in shaping the nebula, as shown in the early work by \citet{beg92}, and confirmed numerically by \citet{van03}. Given that the magnetic field in the pulsar wind is mostly toroidal, if no instability arises to destroy the axisymmetric structure \citep{beg98}, the magnetic field will pile up at the edge of the nebula. Due to magnetic stress the pressure  driving the expansion of the thin shell is higher along the rotation axis of the pulsar than at the equator. Despite the change in boundary pressure, results show that the simple power law expansion rate, derived in the self-similar spherical model still holds. The elongation of the nebula increases with increasing magnetization, so that in principle can be used to constrain properties of the wind. Time dependent results show also that the elongation tends to saturate in a few hundreds years, a period smaller than the age of observed nebulae, and that the saturation time is shorter for smaller magnetization \citep{van03,ldz04}. Interestingly, simulations done by \citet{ldz04} give results  consistent with \citet{van03}, despite the fact that  in the former the pulsar wind is highly anisotropic, and the energy flux is concentrated in the equatorial plane. The reason is that even if the wind is anisotropic, and the post-shock flow has high velocities, in the bulk of the nebula the flow is subsonic and pressure equilibrium holds. This confirms that while wind anisotropy might be important in the modeling of the inner X-ray emitting region,  the global evolution and large scale properties of the nebula depend mostly on the integrated pulsar luminosity.


\section{The inner structure of PWNe}
\label{sec:inner}

\begin{figure}
\label{fig:3}
\resizebox{\hsize}{!}{
\includegraphics[scale=0.50, clip=true]{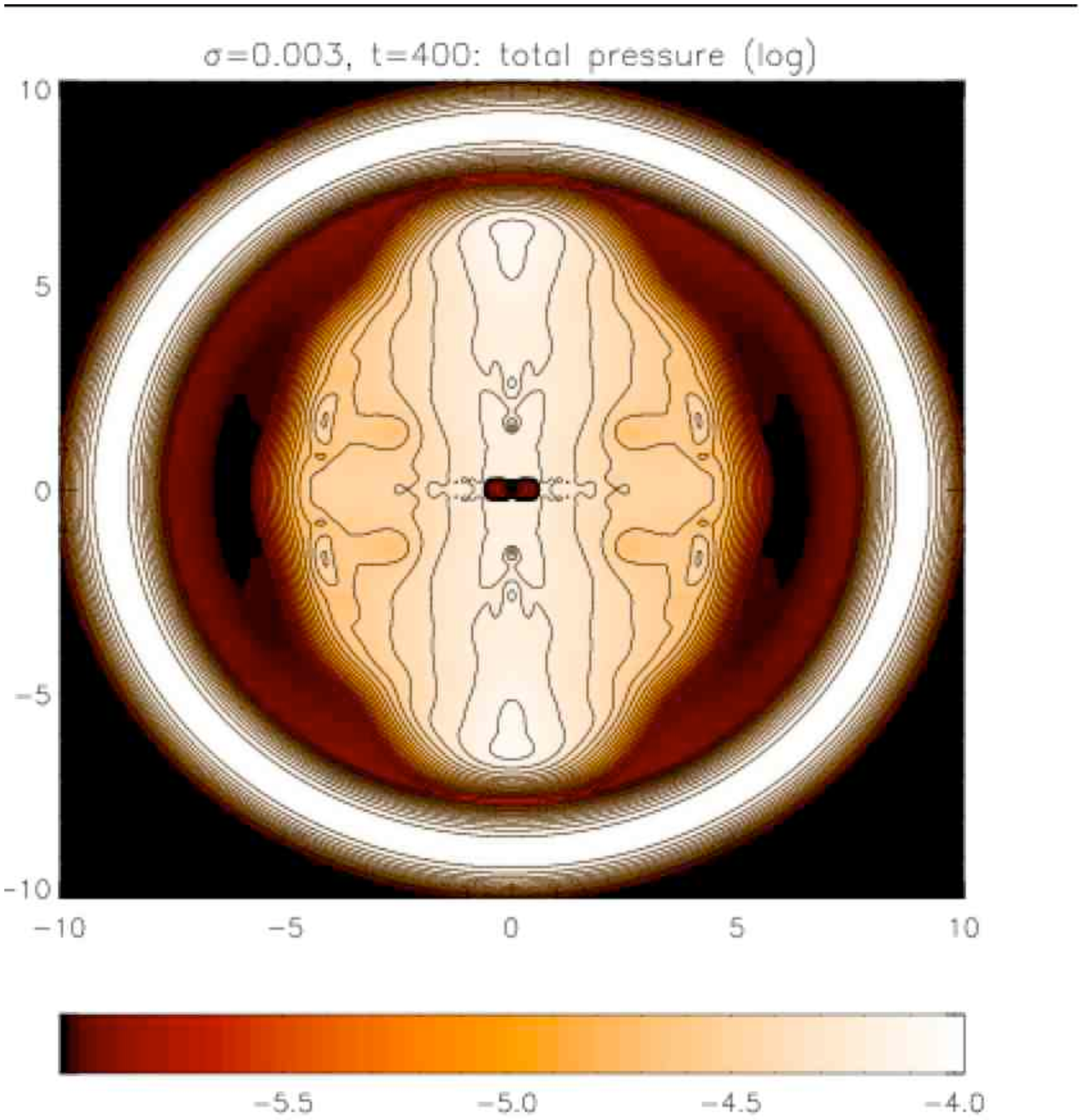}\hspace{0.5truecm}
\includegraphics[scale=0.50, clip=true]{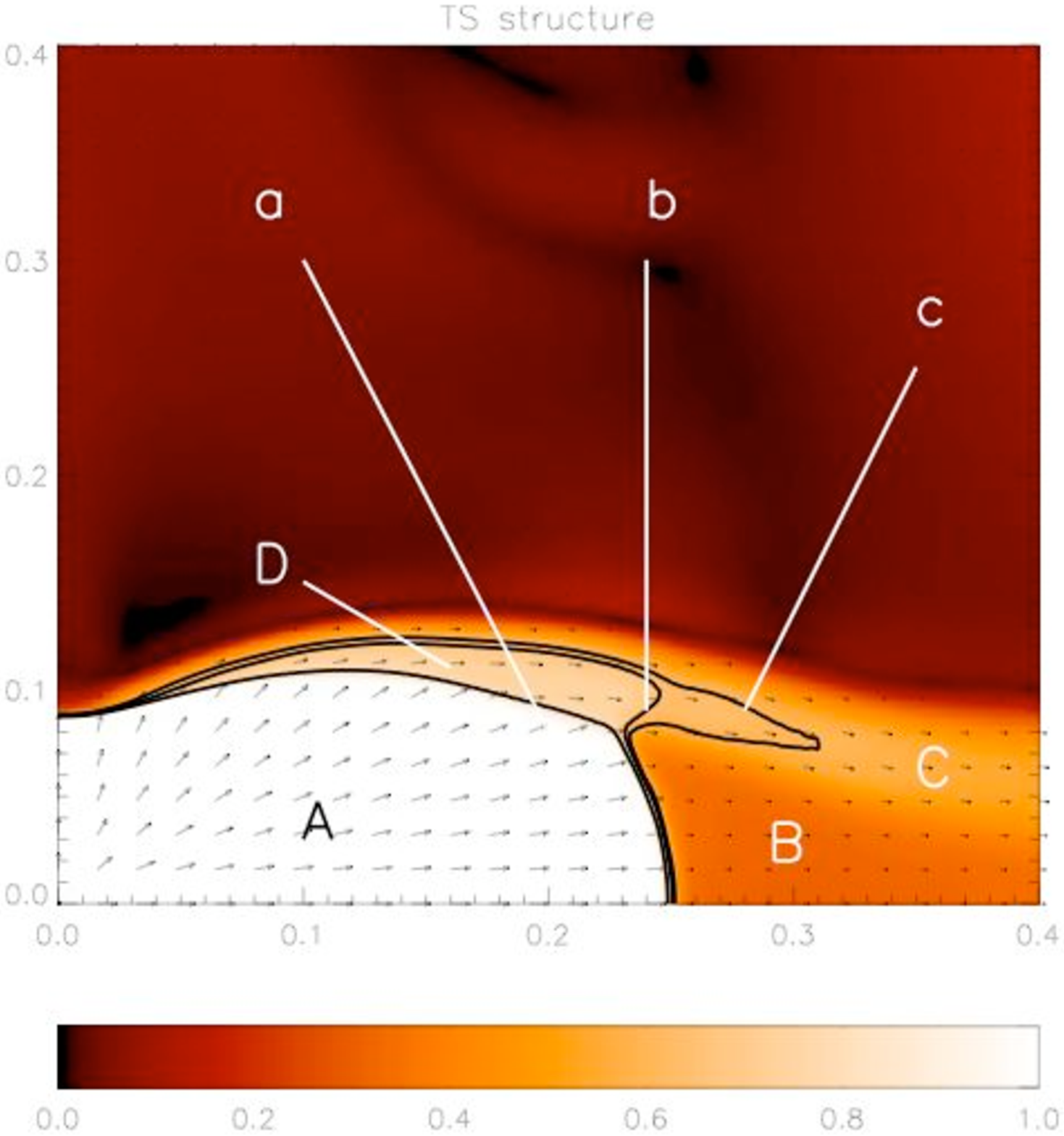}}
\caption{Left panel: global structure of a PWN, pressure distribution (from \citet{ldz04}). Despite the higher equatorial energy flux, the pressure on large scales depends on the cylindrical radius as expected from pressure equilibrium. The higher polar pressure is responsible for the elongation of the nebula. Right panel: structure of the flow at the termination shock, velocity (from \citet{ldz04}). The various labels indicates: A- the relativistic pulsar wind, B- the subsonic equatorial outflow, C- supersonic equatorial funnel, D- super-fastmagnetosonic flow, a- termination shock, b- rim shock, c- fastmagnetosonic surface.}
\end{figure}

Let us now turn our attention to the inner region of PWNe, where the X-ray emission comes from. Simplified 1D models, assume an isotropic energy flux, however, a series of recent numerical simulations both in the force-free \citep{con99,gru05} and in the MHD \citep{bog01,kom06,me06} regimes have shown that, far away from the light cylinder, the flow of an aligned rotator closely resembles the exact monopole solution by \citet{mic73}. This solution predicts that at the typical distance of the TS, the energy flux in the wind has a strong latitudinal dependence of the form $L(\theta)=L_o(1+\alpha\sin^2{(\theta)})$, where $\alpha$ is a measure of the pole-equator anisotropy ($\alpha >>1$), while the magnetic field in the wind $B(\theta)\propto \sin{(\theta)}$. In this case the magnetization in the wind is a function of the polar angle $\sigma(\theta)$. Various numerical simulations of the interaction of such wind with the SNR ejecta have been presented \citep{kom04,ldz04,bog05,ldz06}. Interestingly these results show that, as long as the energy flux is the same, the post shock flow does not depend on the specific values of Lorentz factor or density distribution. In this sense flow dynamics cannot be used to constrain the value of the wind Lorentz factor or composition. In Fig.~\ref{fig:3} the shape of the TS is shown. Due to the higher equatorial energy flux the TS is closer to the pulsar along the polar axis than on the equator. In the low magnetization case ($\sigma_{equator}<0.01$) the shape of the shock can be reasonably reproduced by assuming pressure equilibrium between the wind and the nebula. However, due to magnetic stresses, the nebular pressure is higher on the axis, and the shape of the termination shock becomes more oblate at higher $\sigma$. Due to the TS shape, the flow is slowed down to speed $\sim c/3$ close to the equator, but at higher latitudes, where the shock is oblique, the post shock flow is still super-fastmagnetosonic. Moreover, plasma at higher latitudes is deflected toward the equatorial plane. This flow is partly slowed in the so called ``rim-shock'', however its velocities are still supersonic. As shown in Fig.~\ref{fig:3} the result of the anisotropic energy distribution is that almost all of the downstream plasma is confined in the equatorial region. The inner part  moves with speed $\sim c/3$, while the outer channels has velocity $\sim 0.5c$, in agreement with the value inferred from the difference in the luminosity of front and back side of the torus in the Crab Nebula \citep{shi03}. Numerical simulations seem to give contrasting results regarding the stability of the TS. In some cases the shape of the TS evolves in a self similar manner \citep{vol06,ldz04}, while in other cases it shows significant fluctuations \citep{kom04}. Perhaps such behavior might be associated with boundary conditions on the expansion velocity and the average magnetization in the nebula, but no detailed study has yet been presented on how the boundary velocity affects the overall inner flow pattern.

As the flow expands in this equatorial sheet away from the TS, the magnetization increases, until equipartition is reached. Given that the magnetization in the wind is higher in the equatorial region, the flow reaches equipartition close to the equator, before that at higher latitudes. The post-shock flow is only mildly relativistic, so hoop stresses can produce collimation. Given the extra degree of freedom with respect to  the simple radial model by KC84 (the plasma can now move also in then $\theta$ direction), once equipartition is reached, the magnetic pressure prevents further compression of the magnetic field, and the flow is diverted  back toward the axis. This is the process that causes collimation and the formation of a jet along the axis itself. Fig.~4 shows numerical results for the magnetization and velocity profile inside a nebula. Obviously the values of the wind magnetization play a key role in the formation of the jet. If magnetization is too small, then equipartition is not reached inside the nebula, hoop stresses never become effective, the equatorial channel survives to the edge of the PWN and no jet is formed. At higher magnetization  equipartition is reached in the close vicinity of the TS, and most of the plasma is diverted and collimated in a jet. The size and flow velocity in the jet are a function of $\sigma$. For values $\sigma\sim 0.03$ the plasma speed in the jet is $\sim 0.7 c$ in agreement with observation of the jet in Crab Nebula \citep{wei00,hes02,mel05}. Simulations also show that there is a global meridional circulation inside the nebula associated with the jet formation , with typical speeds $\sim 0.1 c$. Even if this speed is quite subsonic (as previously mentioned in the context of global pressure distribution), it might have important consequences for the interaction with the SNR swept-up shell, and especially on the development of local shear instabilities and on the possible mixing with cold ions \citep{lyu03}.

\begin{figure}
\label{fig:4}
\resizebox{\hsize}{!}{
\includegraphics[scale=0.5, clip=true]{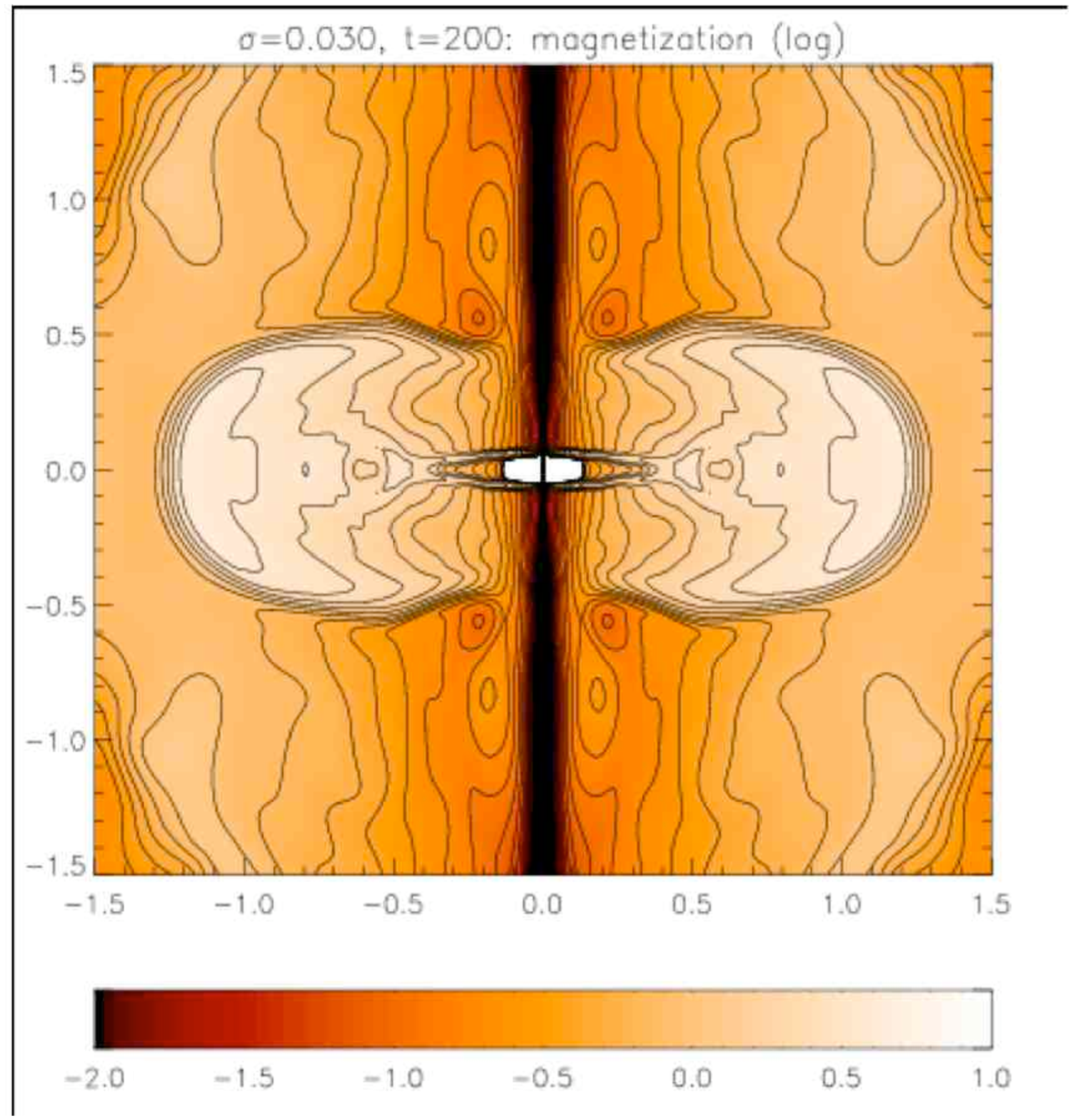}\hspace{0.5truecm}
\includegraphics[scale=0.5, clip=true]{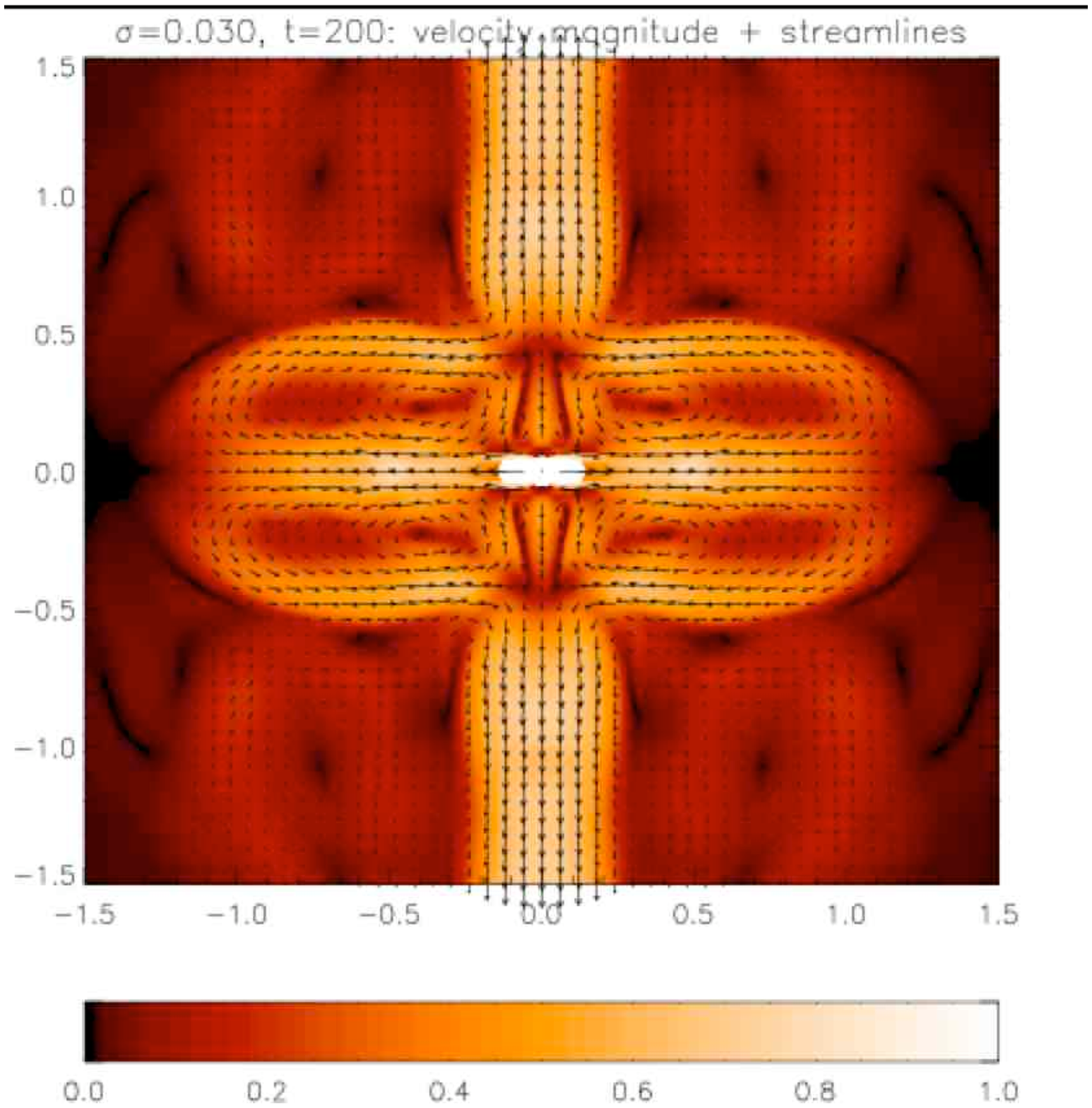}}
\caption{Result of a numerical simulation showing the details of the formation of the polar jet (from \citet{ldz04}). Left panel: magnetization inside the nebula. Right panel: flow velocity. Notice that when equipartition is reached the flow is diverted and collimated toward the axis.}
\end{figure}

The picture presented above can be modified to take into account the fact that a pulsar, most probably, is an oblique rotator. It has been shown both theoretically \citep{bog01b} and numerically \citep{spi06}, that even in the case of an oblique rotator, the energy distribution in the wind is essentially the same as the aligned case (in the case of split monopole it can be shown to be identical). However, the presence of a folded current sheet that might extend to higher latitudes, has important consequences on the magnetization. If the alternating  magnetic field in this striped wind region is dissipated, and this can happen either in the wind \citep{lyu01b,kir03} or at the termination shock itself \citep{lyu03b,lyu05}, then the magnetization will drop to zero on the equator and will reach a maximum at intermediate latitudes, depending on the obliquity of the pulsar. Given that, in MHD codes, the TS is a thin discontinuity, there is no difference between dissipation in the wind or at the shock, and one can chose the former for numerical simplicity. We need however to remember that, in simulations of aligned rotators, the basic assumtpio is that the equatorial current sheet survives in the nebula. In reality in the shocked plasma the current sheet can undergo instability and dissipation, the details of which have not being investigated.

The presence of an unmagnetized region close to the equator, adds complexity to the picture presented above. In particular, now the equatorial flow does not reach equipartition inside the nebula, and an equatorial channel survives to much larger distances from the termination shock. However, at higher latitudes, where dissipation is marginal, and the magnetization is higher, equipartition can be reached and the flow is then diverted to the axis and collimated into a jet. Fig.~5 show the flow structure. As will be discussed in more detail in the following, emission maps based on the results of numerical simulations suggest that a striped wind model is more promising in explaining the observes inner-ring outer-torus structure of many PWNe. The extra complexity added by the presence of a striped wind region, can be used as a probe for the properties of the wind. In particular recent results \citep{ldz06} have shown that, given the same average magnetization, a bigger striped wind region, leads to weaker jets, less intense hoop-stresses, and a bigger size of the termination shock, all of which are observable.

\begin{figure}
\label{fig:5}
\resizebox{\hsize}{!}{
\includegraphics[scale=0.5, clip=true]{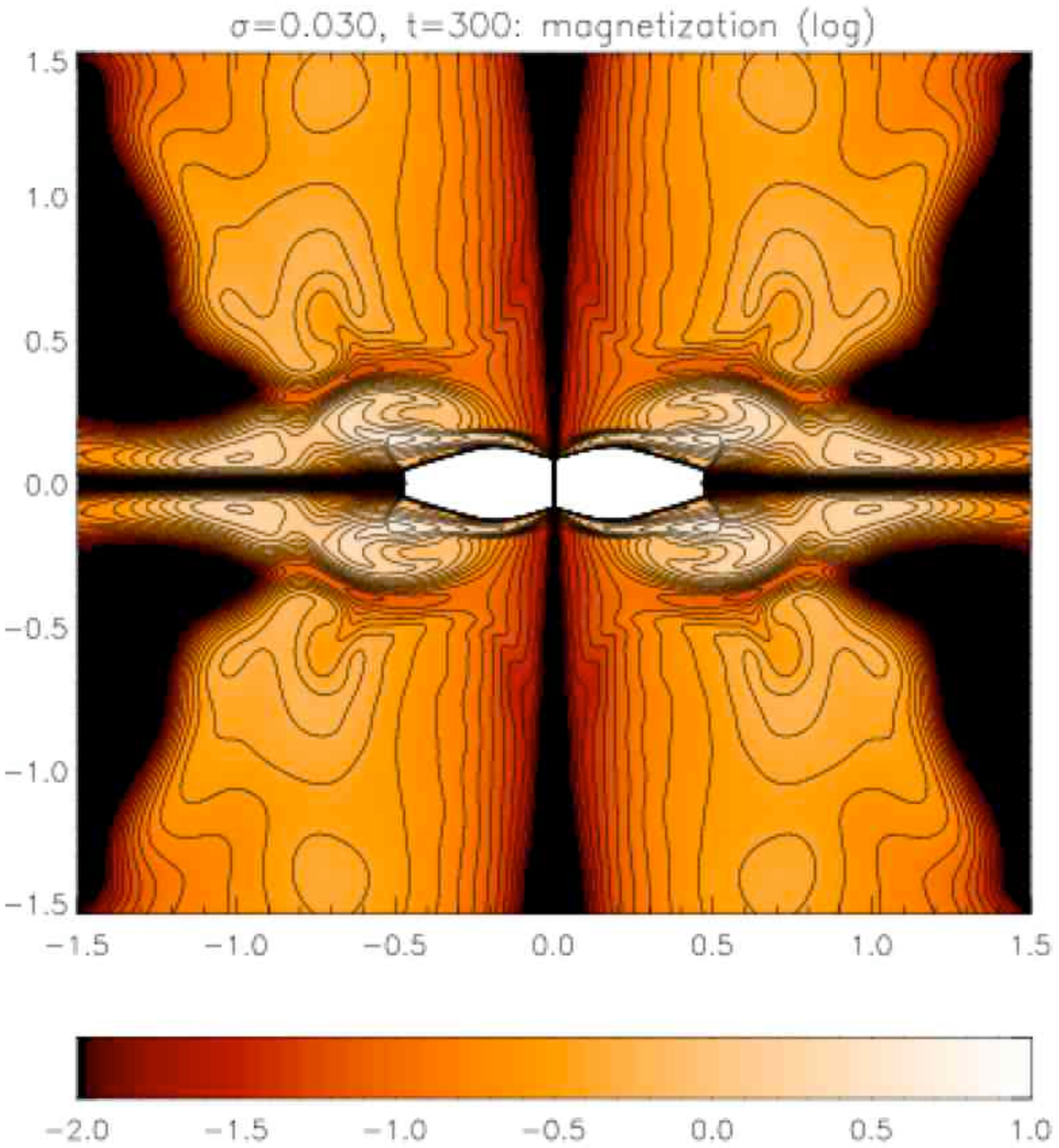}\hspace{0.5truecm}
\includegraphics[scale=0.5, clip=true]{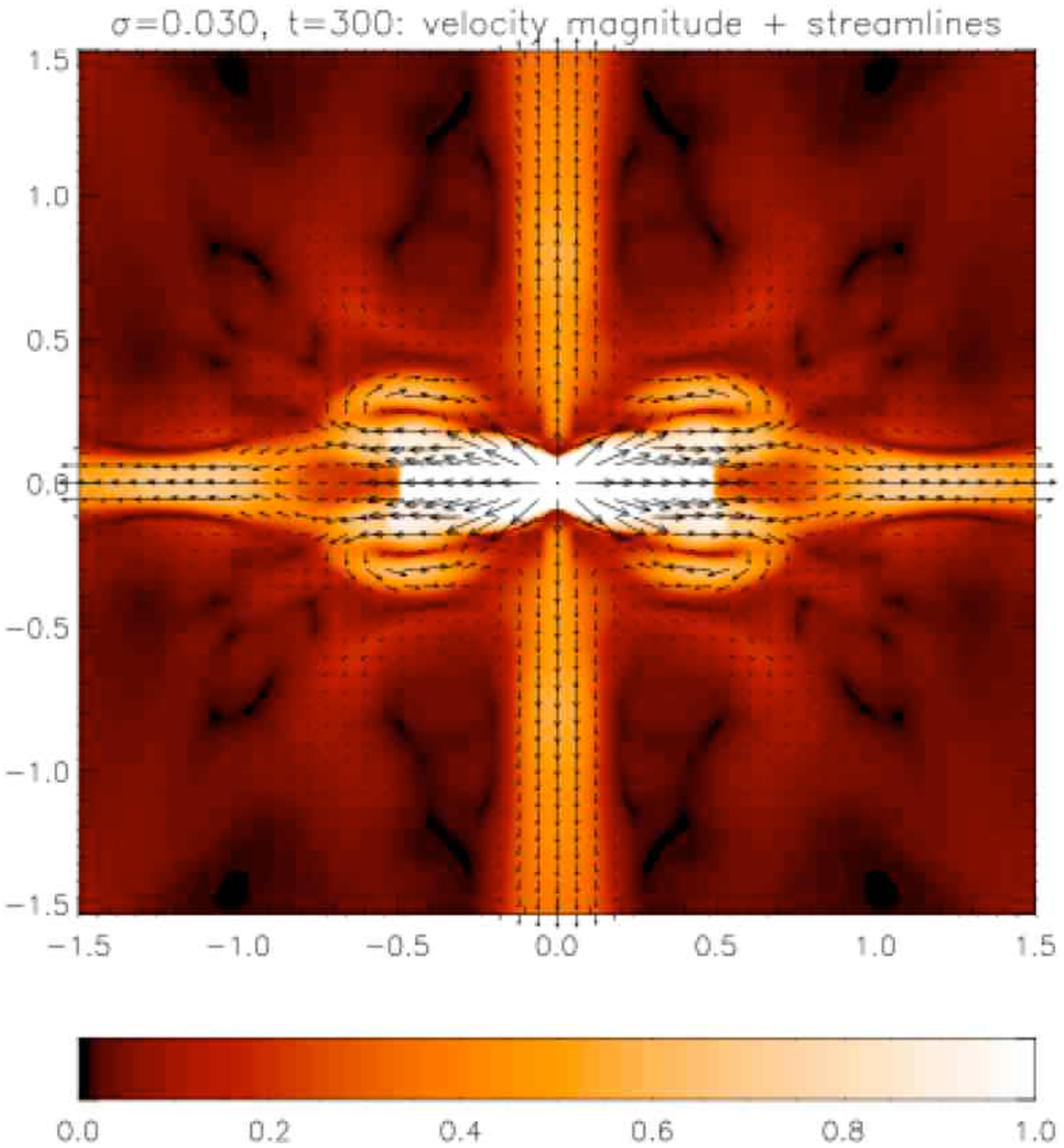}}
\caption{Effects of a striped wind region: result of a numerical simulation showing the detail of the formation of the polar jet (from \citet{ldz04}). Left panel: magnetization inside the nebula. Right panel: flow velocity. An equatorial fast outflow remains, while only material at intermediate latitudes is collimated toward the axis.}
\end{figure}

Even if the fluid structure recovered in relativistic MHD simulations clearly shows the formation of a jet, to properly assess their validity  one must consider the emission and compare results based on simulations with observations. Despite the fact that, variations in the particles' energy distribution and acceleration along the TS are poorely known, to a first order approximation one might assume a power-law injection distribution, and an uniform efficiency. Under this assumption, to obtain X-ray maps, one must then compute the adiabatic and synchrotron losses. In \citet{kom04} this was done by assuming an exponentially decaying factor, with spatial constant cooling time depending on the frequency of observation and an average magnetic field. A more sophisticated strategy was adopted by \citet{shi03} where several particles energies are evolved independently along streamlines. While the latter is probably the more accurate, it can be applied only in steady state cases, where the streamlines can be traced without ambiguities. An intermediate approach was proposed by \citet{ldz06}, where the maximum energy $\epsilon_{max}$ of the particles' distribution was evolved with the flow, accounting both for synchrotron and adiabatic losses. In this case the particle distribution remains a power low with a break at $\epsilon_{max}$.

\begin{figure}
\label{fig:6}
\resizebox{\hsize}{!}{
\includegraphics[angle=-90,scale=3,clip=true]{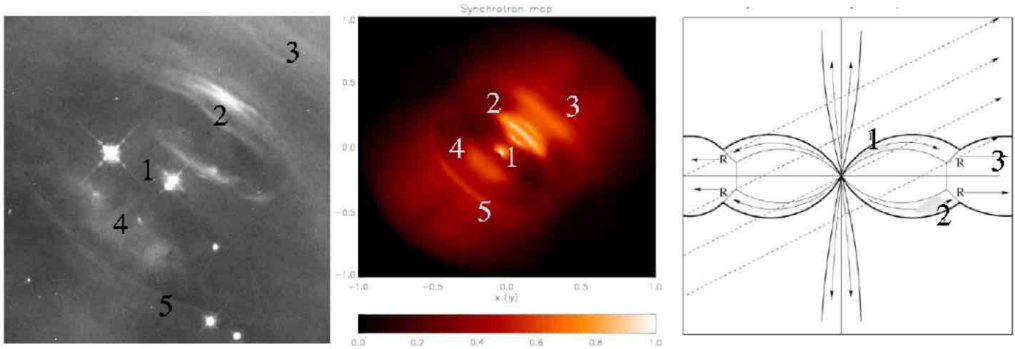}}
\caption{Emission features. Left panel: optical HST image of the inner region in the Crab Nebula \citep{hes95}. Central panel: optical synchrotron map based on a numerical simulation of the flow inside a PWN, in the case of a striped wind. Schematic interpretation of the origin of some of the observed feature \citep{kom04}. The various labels indicate: 1- the knot, 2- the wisps, 3- the main torus, 4- the anvil, 5- the back side of the wisps. Notice that this features are recovered in the simulated map.}
\end{figure}

Preliminary results show that it is possible to reproduce both the inner ring (or wisps region) and the outer torus, only by assuming the presence of a striped wind region. Indeed in the case of no striped wind, maps usually produce a single luminous arch. In Fig.~6, we show an optical image of the Crab Nebula, a map based on a simulation with striped wind (in this case, for comparison in the visible band, synchrotron cooling is neglected), and a schematic picture from \citet{kom04}, to illustrate the origins of the various features. The knot and the inner ring are both explained as due to the high velocity flow in the immediate post shock region, at intermediate latitudes. However, by looking at Fig.~5, we see that, in the striped wind case, a fast equatorial channel survives, extending far from the TS. It is the emission coming from this channel that is responsible for the main torus observed in the Crab Nebula. Recent investigations \citep{vol06,ldz06} have shown that the size and shape of the striped wind region, have important observational consequences. In general a larger unmagnetized sector will produce a more structured inner emission; the region corresponding to the inner ring might split in several minor arches; the extent of the X-ray nebula is bigger; the jet is thinner and less enhanced. If the magnetization is not high enough, energetic particles can survive all along the jet and accumulate in a mushroom cap at the edge of the nebula.  Simulations also seem to disfavor configuration with a higher magnetization close to the pole \citep{aro98}. In fact in this case, hoop stresses are less effective and no observable bright extended  jet is formed. While modeling of the equatorial region does not need any accurate treatment of the synchrotron losses (the feature are observed also in radio \citep{bie04}), these are fundamental in understanding the jet, and simulated map clearly show that the jet is not visible over the background, in optical or radio. 

\begin{figure}
\label{fig:7}
\resizebox{\hsize}{!}{
\includegraphics[angle=-90,scale=3,clip=true]{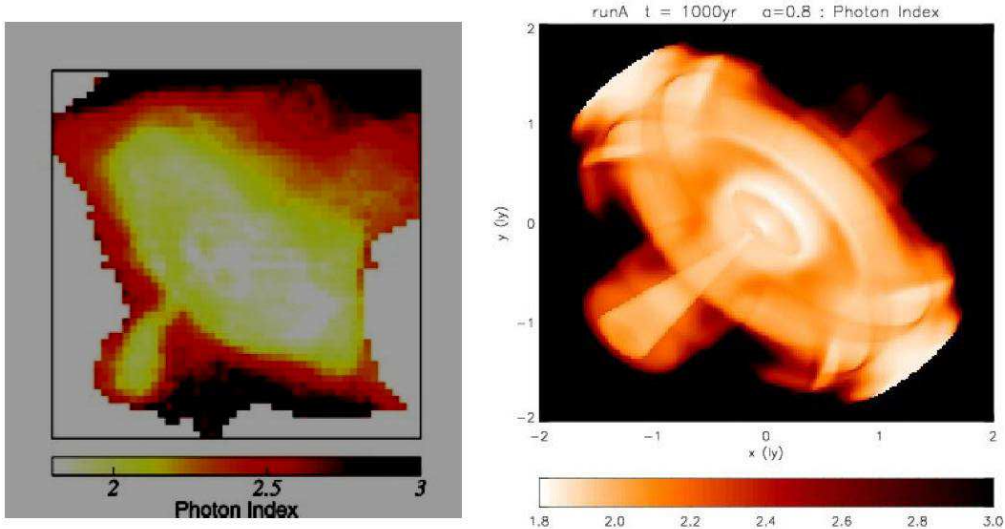}}
\caption{Spectral properties of PWNe. Left panel: map of the X-ray photon index in the Crab nebula \citep{mor04}. Right panel: photon index computed from a numerical simulation. Dissipation was added in the polar region to reproduce the correct value for the jet.}
\end{figure}

Interestingly, a detailed study of emission with spectral maps, polarization properties, and integrated spectra, can also be used to better constrain the existing models. If the particle energy distribution is somehow followed along a streamline, it is possible to reconstruct spectral maps. In Fig.~7 an X-ray map of the spectral index based on simulations is presented together with a map of the Crab Nebula \citep{mor04}. The main features are indeed recovered. We also see that simulations produce maps where the spectrum appears to flatten moving toward the main torus, without the need to assume any re-acceleration. This is just due to the Doppler boosting effect. At low speeds, when observing at a given frequency, we are implicitly sampling particles all at the same energy (let us neglect variation of magnetic field). When velocities are relativistic, the  particle energy responsible for emission at a given frequency, depends also on the Doppler boosting. The spectrum appears to flatten in the torus, because, given the higher speed, we are looking a lower energy particles. However, map based on simulations fail to recover the correct spectrum in the jet: the simulated spectrum is much steeper than what is observed, due to excessive synchrotron losses. To reproduce the observations, one must assume some form of dissipation and re-energization along the axis. This could be associated with local instabilities of the toroidal magnetic field \citep{beg98} which are not captured by axisymmetric simulations, but for which there is much observational evidence \citep{pav03,mor04b,mel05,del06}. By considering the integrated spectra we can obtain some constraints on the magnetization. Preliminary results show that $\sigma>0.03$ is required and a configuration with a smaller striped wind region are favored. Moreover, integrated spectra show a flattening above the CHANDRA band, which has been interpreted as due to the high velocity in the post-shock region. In this case in fact, the size of the nebula at higher energies, does not seem to drop substantially, as it does from optical to X-ray.

\begin{figure}
\label{fig:8}
\resizebox{\hsize}{!}{
\includegraphics[angle=-90,scale=3,clip=true]{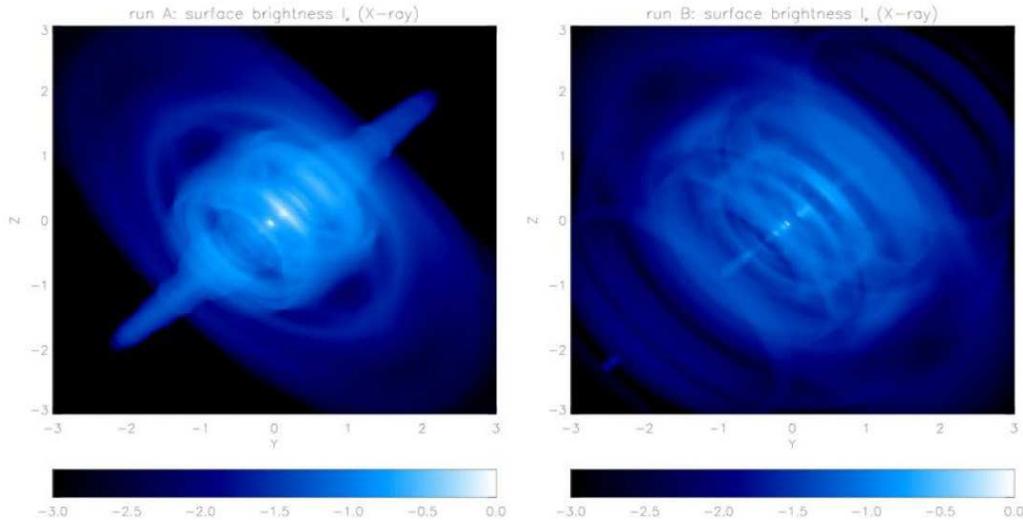}}
\caption{Effect of the striped wind on the X-ray emission features (from \citet{ldz06}). Both maps are based on simulations with the same average wind magnetization, but the one on the left has a much smaller striped wind region. Notice the different extent of the emission, the presence of more arches in the right image, and the luminosity of the jet.}
\end{figure}

Polarization might also be used to derive useful information about the magnetic field structure. Emission maps depend strongly on the flow velocity inside the nebula, but they are not very sensitive to the presence of a small scale disordered component of the magnetic field. It is thus not possible at the moment to use them to properly constrain the geometry of the field. On the contrary, polarization might prove very useful. Given that only optical polarization is available, any study should be limited to the brightest features, like the wisp, inner ring and torus. The effect of flow velocity on the polarization angle has been discussed by \citet{me05b} and \citet{ldz06}. By comparison with old optical polarization maps, one can immediately see that some of the properties are recovered. In \citet{sch79} it is also interesting to note the presence of a region which can be identified with the jet of Crab. More recent data (Graham, private communication) have also shown that the knot has a high degree of polarization, and that the polarization angle is consistent with this feature originating in the post shock flow. 


\section{Time variability} 
\label{sec:time}

Most of the work done in the past has focused on the explanation of the main observed features that appear to be quite persistent on long timescale. However, there is increasing interest in the understanding of time variability. Variability of the wisps in the Crab Nebula has been known since their discovery (see references in \citet{aro98}), and high resolution observations \citep{hes02,bie04} have allowed us to study its behaviour in detail. The jet in Vela also appears to be strongly variable \citep{pav03}, as does the jet of both the Crab \citep{mor04b,mel05} and B1509 \citep{del06}. B1509 shows also variability in the inner ring. There is also evidence for the presence of variable knots (non axisymmetric features) both in Crab and B1509. Variability in the jet, which usually has timescale of years, has been associated with kink or sausage modes in the strong toroidal field, or even with the fire-hose instability \citep{tru88} in the case of Vela. Recently in B1509, variable knots have been observed inside the TS, on the side of the counter-jet \citep{del06}. These are non axisymmetric features, however it has been speculated that they might be the signature of instability in the TS cusps at the base of the counter-jet.

More interesting is the case of the wisps, which show variability on a timescale of weeks. The presence of high velocity flow channels inside the nebula suggests the possibility of Kelvin-Helmholtz instability \citep{beg99}, however it seems that it cannot account for the luminosity variations required in the wisps (the fluctuations  seem not to exceed 10\%, \citet{me06b}). However, the flow inside the nebula is strongly dynamic, and variations in the meridional circulation pattern, can provide a possible explanation. In simulations by \citet{bog05} the  synchrotron luminosity varies on short time scales (few months). Results by \citet{kom04} also suggest that the flow might have a feedback action on the TS, causing it to change shape, thus inducing a change in the wisps, which originate in the post-shock flow. 

A different model for the variability of the wisps has been proposed by \citet{spi04}. The model is based on the assumption that high energy ions are present in the equatorial region of the wind. There are several reasons for this claim, most important the fact that, in kinetic simulations of acceleration in a strong shock, a pure pair plasma does not produce a power-law distribution \citep{ama06}. Given that ions have a much larger larmor radius, of order of the size of the wisp region, they introduce a substantial deviation from a pure fluid picture. In particular, electrons are compressed by the gyration of ions and the emission is enhanced. The model is able to reproduce the observed time scale variability and the average distance between the torus and inner wisps. However, these results are based on a 1D spherically symmetric model, do not take into account the energy flux anisotropy, and the shape of the TS, and assume coherent motion of the ions in an equatorial sector. The ion cyclotron and MHD model are by no mean incompatible. If  the presence of high energy ions is confirmed by observations \citep{ama03,hor06}, an hybrid approach that integrates both features would be necessary for better models. A third model for the wisps dynamics has been presented by \citet{foy06}, based on the idea of synchrotron cooling instability. However, in order to work, this model assumes most of the energy in the nebula to be carried by the X-ray emitting particles, in contrast with the standard assumption that lower energy particles dominated the bulk flow.  


\section{Conclusion}
\label{sec:concl}

In the last few years, thanks to a combination of high resolution observations, and numerical simulation, our understanding of the evolution and internal dynamics of PWNe has greatly improved. We have been able to reproduce the observed jet-torus structure and to clarify how the jet is formed in the post shock flow, as a function of the wind magnetization. Simulated maps can reproduce many of the observed features, and the correct behavior of the spectral index in X-ray. Results suggest that, the best agreement is achieved in the case of a striped wind, even if MHD simulations are not able to distinguish between dissipation of the current sheet in the wind or at the TS. It appears that the size of the striped wind zone affects both the complexity of the observed structure and the intensity of the jet. This has open the possibility of using X-ray imaging to constrain the pulsar wind properties; already the rings and tori observed in many PWNe have been used to determine the spin axis of the pulsar \citep{rom05}. Interestingly in the Crab the inner ring does not appear boosted, while the wisps (which are interpreted as its optical counterpart) are. 

There are still however unsolved questions, and possible future developments for research in this field. The interaction of the relativistic hot plasma with the confining SNR, has not been considered in detail, in the framework of the new interior models. It is not clear how the Rayleigh-Taylor instability is affected by the global circulation, and what kind of feedback is exerted by the filamentary network on the inner flow. All simulations are axisymmetric, thus none is able to address the problem of the stability of the toroidal field. Observations on the other hand show clear non axisymmetric variability in the jet. Such instability might be important in modeling the jet emission, especially given that present results fail to reproduce the correct synchrotron spectrum. It is not clear if small scale disordered field is present in the inner region (may be a residual of the dissipation in the TS of the striped wind). A combination of simulations and polarimetry might help solving this question. Finally it should be borne in mind the possible  presence of ions and the modifications they will produce.

\section*{Acknowledgments}
This work has been supported by NSF grant AST-0507813, and by NASA grant NAG5-12031. NB was supported partly by NASA through Hubble Fellowship grant HST-HF-01193.01-A, awarded by the Space Telescope Science Institute, which is operated by the Association of Universities for Research in Astronomy, Inc., for NASA, under contract NAS 5-26555. 




\end{document}